\documentclass[trackchanges,twocolumn,twocolappendix]{aastex701}
\usepackage{amsmath}

\begin{document}

\title{Large Scale Spectrophotometric Relative Flux Calibration for the Roman High Latitude Wide Area Survey}

\author[orcid=0009-0005-9574-7792]{Alan B. H. Nguyen}
\affiliation{Department of Physics, Yale University, New Haven, CT 06511, USA}
\email[show]{alan.nguyen@yale.edu}  

\author[]{Gregory Walth}  
\affiliation{IPAC, California Institute of Technology, Pasadena, CA 91125, USA}
\email{gwalth@ipac.caltech.edu}  

\author[]{Ashley J. Ross} 
\affiliation{Department of Astronomy, The Ohio State University, Columbus, OH 43210, USA}
\affiliation{Center for Cosmology and AstroParticle Physics, The Ohio State University, Columbus, OH 43210, USA}
\email{ashley.jacob.ross@gmail.com}  

\author[]{James Colbert} 
\affiliation{IPAC, California Institute of Technology, Pasadena, CA 91125, USA}
\email{colbert@ipac.caltech.edu}  

\author[]{Jaide Swanson} 
\affiliation{Department of Physics \& Astronomy, Ohio University, Athens, OH 45701, USA}
\email{js956022@ohio.edu}  

\author[orcid=0000-0002-2885-8602]{Nikhil Padmanabhan} 
\affiliation{Department of Physics, Yale University, New Haven, CT 06511, USA}
\affiliation{Department of Astronomy, Yale University, New Haven, CT 06511, USA}
\email{nikhil.padmanabhan@yale.edu}  

\author[]{Yun Wang} 
\affiliation{IPAC, California Institute of Technology, Pasadena, CA 91125, USA}
\email{wang@ipac.caltech.edu}

\begin{abstract}

We consider the application of a \textit{{\"u}bercalibration}-like relative flux calibration to the grism observations of the Roman High Latitude Wide Area Survey (HLWAS). We propose a simplified model of the calibration with an independent flat field for each detector in each exposure of the focal plane. In addition, we include two wavelength dependent components: a single wavelength throughput curve, modulated by a simple parabolic model for the throughput as a function of a source's focal plane position. We consider the impact of the dither scale, as well as the calibrator magnitude cuts. We show that the width of the calibration residuals can be reduced to less than 1.5 mmag, or 0.15$\%$ in flux, within the optimal dither range 50-240$^{\prime\prime}$. This wide range allows for significant flexibility in optimising other parts of the observing program without diminishing the effectiveness of the relative flux calibration. We also discuss some improvements to the methodology that must be strongly considered before the calibration can be applied to real data. Finally, although we focused on spectroscopic component of the HLWAS here, our formalism and results should carry over to the imaging surveys as well.

\end{abstract}



\section{Introduction} \label{sec:intro}

A challenge for astrophysics experiments is relating the detector signal to the underlying physical quantity. An astronomical camera collects counts in each pixel, a quantity that is proportional to the number of incident photons. This output must then be calibrated to yield physical flux densities. The key scientific programs of next generation surveys will demand even more precise spectrophotometric calibrations to carry out their goals. For example, the galaxy clustering measurements from the current DESI and Euclid missions rely on accurate photometric calibration, as the clustering signal is a varying function of scale and sky position \citep{laureijs_euclid_2011, collaboration_desi_2016}. These signals are functions of the fractional matter overdensity and can be easily overwhelmed by small percentage level systematic errors due to errors in calibration.

The Nancy Grace Roman Space Telescope (Roman) is NASA's next flagship mission, set to launch by 2027 \citep{spergel_wide-field_2013}. In this paper we will focus on calibration techniques relevant to Roman's third core community survey: the High Latitude Wide Area Survey (HLWAS). The HLWAS is designed to help determine the cause of the accelerated expansion of the universe and if the culprit is indeed a new energy component, like dark energy (DE), then how does it behave over time? Roman will accomplish this by tracking both the expansion history of the universe as well as the growth history of cosmic structure, allowing for precise testing of possible models of the expansion including DE and modifications to Einstein's general relativity. 

As part of the HLWAS, Roman will make use of grism-based slitless spectroscopy to disperse light onto its focal plane. Hereafter referred to as the High Latitude Spectroscopic Survey (HLSS). The grism has a near-infrared (NIR) wavelength coverage of 1.00--1.93 microns, and disperses light onto the 0.28 deg$^2$ Wide Field Instrument (WFI). In this wavelength range, Roman is sensitive to about 10 million H$\alpha$ emission line galaxies from z = 1--2 and 2 million [O III]$\lambda$5007 emission line galaxies from z = 2--3. These dense, large volume observations will enable robust dark energy constraints using baryon acoustic oscillation (BAO) and redshift space distortion (RSD) measurements from Roman alone \citep{wang_high_2022}. While this survey will be prone to systematics like spectral confusion and redshift interlopers, the effect we focus on will be the time and wavelength varying calibration of the spectra and photometry. 

The requirement for the relative spectrophotometric flux calibration is two percent, with a goal of one percent. While the absolute flux calibration may only impact the total number of galaxies, the density fluctuations much of cosmology is interested in will be unaffected.  As the HLSS will be flux limited at $1.5\times 10^{-16}$ erg s$^{-1}$ cm$^{-2}$, inaccuracies in the calibration across the survey footprint will induce artificial density fluctuations \citep{wang_high_2022}.

A standard calibration approach is to simply compare observations to known standards. This is ideal for fixing absolute calibrations of exposures, but suffers from requiring many known standards across a plethora of desired filters. The alternative is to use self-calibration, where repeated observations of the same objects are used to determine the relative calibration between exposures, then leaving the issue of absolute calibration to the comparison with some known standards. As the intrinsic brightness of astronomical calibrators is constant between exposures, they may be marginalised over to extract the relative calibrations. This self-calibration technique was explored previously in several papers \citep{padmanabhan_improved_2008, panstarrs_hyper_2016, markovic_large-scale_2017}. We will expand on this work and develop a basic framework for a full spectrophotometric calibration pipeline, allowing Roman and other future surveys to calibrate the detector response as a function of position and wavelength.

We present a spectrophotometric relative flux calibration in two steps, returning calibrations as a function of detector exposure, wavelength, and a source's focal plane position. We first consider calibrations in fixed wavelength bins, minimising an effective $\chi^2$ to obtain flat field calibrations for each detector exposure as well as a smoothly varying throughput as a function of distance of the WFI centre. Then, having applied the first set of calibrations to the measurements, we find a set of wavelength dependent throughputs using a similar effective $\chi^2$ to complete the full spectrophotometric calibration. This two step process helps to break the degeneracy between the latter two sets of calibrations, improving the power of the calibration as well as the speed of the convergence.

Our paper is structured as follows. In Section~\ref{sec:simoutline} we outline the simulation we use to test our spectrophotometric calibration technique and show the posterior distribution we use to find the optimal calibration functions in Section~\ref{sec:likemax}. In Sec.\ref{sec:ubercalresults}, we apply our technique to a simulation of the Roman HLSS. 
Some extensions to the work are discussed in Sec.\ref{sec:discussion}.

\section{Simulation Outline}
\label{sec:simoutline}

\subsection{The Roman HLSS}
\label{sec:hlwas}

\begin{figure}[ht!]
\plotone{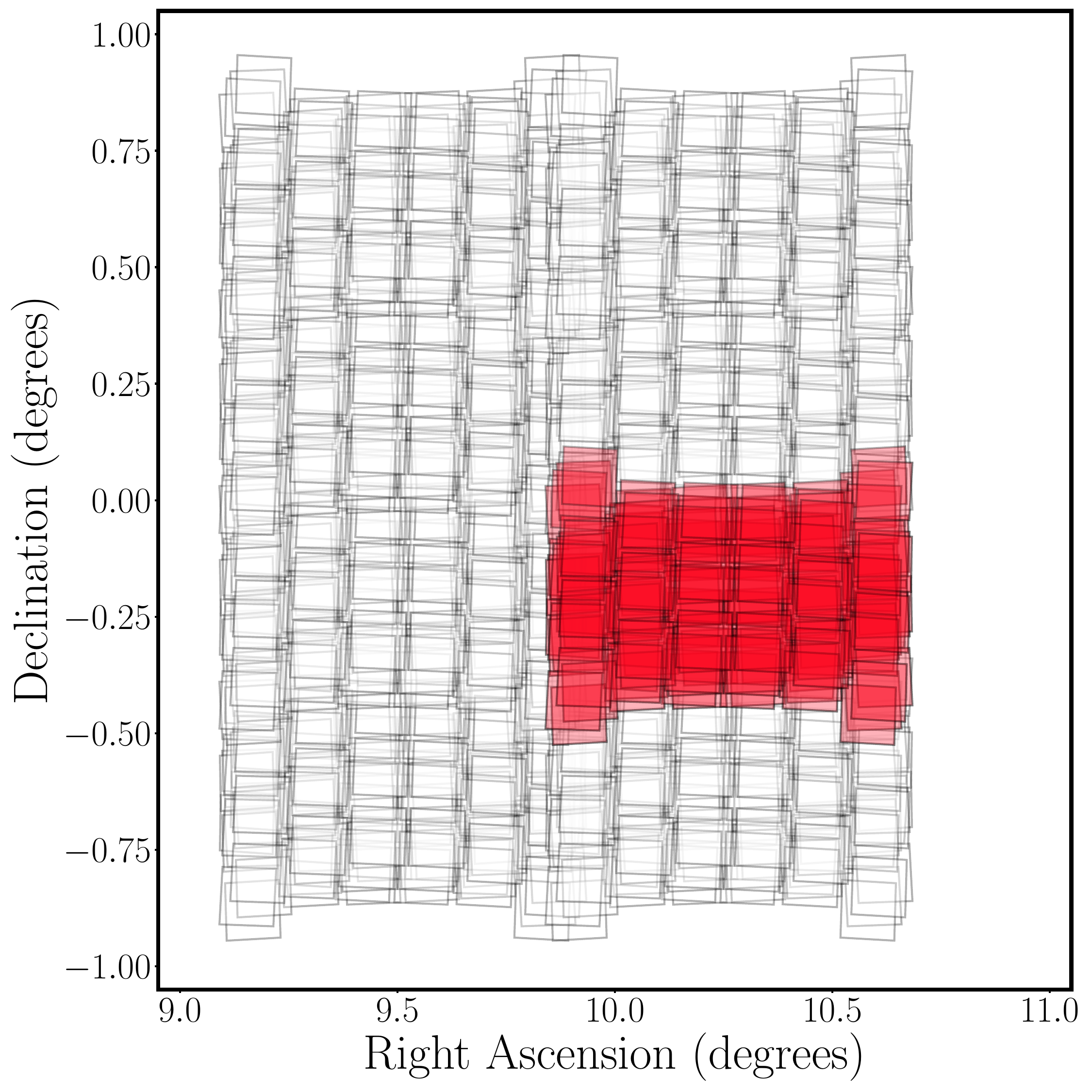}
\caption{We plot the detector outlines for our reference survey strategy with a dither $d_{\rm RA} = 90^{\prime\prime}$. We highlight a single ``chunk'' in red in the right column of observations.
\label{fig:examplesurvey}}
\end{figure}

The Roman HLSS will use a grism to perform slitless NIR spectroscopy, obtaining redshift information from distant galaxies in the wavelength range 1.00-1.93 $\mu$m. The focal plane is comprised of an array of 18 H4RG-10 Sensor Chip Assemblies (SCAs) or detectors, making up a field of view of approximately 0.4 by 0.8 degrees (0.281 square degrees accounting for chip gaps). Each detector is a square made up of 4096 pixels on a side.\footnote{4088 pixels after removing reference pixels.} The nominal strategy for the Roman HLSS grism observations consists of eight exposures around a given pointing, using four
roll angles and two dithers. We denote these eight exposures as a "chunk", with an example highlighted in Figure~\ref{fig:examplesurvey}.


We have set up several simulated Roman-like surveys to test the efficacy of various survey strategies on the relative flux calibration. In the remainder of this section, we describe the simulated survey setup.

\subsection{Simulating the calibrator observations}
\label{sec:obs_sim}

Starting with the true magnitude of the calibrator split into 10 wavelength bins ($m_{j\lambda}^*$), we include three effects. We apply a constant magnitude offset or flat field calibration for each detector exposure ($k_i$), an overall wavelength dependent response ($r_{\lambda}$), and a smoothly varying field of view (FoV) throughput ($\phi_{j\lambda}$) to construct the measurement matrix $m_{ij\lambda}$ which is indexed by the $i^{\rm th}$ detector exposure, $j^{\rm th}$ calibrator, and $\lambda^{\rm th}$ wavelength bin. We assume the shape of the spectrum is flat. Working in magnitudes, we have:
\begin{equation}
    m_{ij\lambda} = m_{j\lambda}^* - k_i - r_{\lambda} - \phi_{j\lambda} + e_{j\lambda},
\end{equation}

\noindent
where $m_{j\lambda}^*$ is the integrated magnitude over the wavelength range 1.00-1.93 $\mu$m, and $e_{j\lambda}$ is the Poisson error with width $\sigma_{j\lambda}$.

The set of $k_i$ is drawn from a Gaussian distribution centred at 0 with a width $\sigma_k$ = 0.5. For the FoV throughput, we adopt a simple paraboloid model (see Section~\ref{sec:focalplane}):

\begin{equation}
    \phi_{\lambda}(d_j) = -2.5\log_{10}\left(A_{\lambda} d_{j\lambda}^2 + 1\right) = \phi_{j\lambda},
\end{equation}

\noindent
where we denote the distance from the WFI centre in degrees of the $\lambda^{\rm th}$ wavelength bin of the $j^{\rm th}$ calibrator as $d_{j\lambda}$. Note that it is the $A_\lambda$ values that we will recover. We describe and show the constant throughput curve ($r_\lambda$) in Section~\ref{sec:focalplane}.

Thus we write the residual miscalibrations as $\varepsilon_{ij\lambda}$:
\begin{equation}
    \varepsilon_{ij\lambda} = \hat{k} + \hat{r}_{\lambda} + \hat{\phi}_{j\lambda} - k_i - r_\lambda - \phi_{j\lambda},
    \label{eq:residuals}
\end{equation}

\noindent
where the hat $\ \hat{} \ $ denotes an estimated quantity. 

Finally, we adopt a similar set of assumptions as \cite{markovic_large-scale_2017}: the calibrator brightnesses are independent, the exposure calibrations are independent, the wavelength bin calibrations are independent and there is no cross covariance between any of these quantities. This allows for a relatively simple form of the likelihood.

\subsection{Variation across the focal plane}
\label{sec:focalplane}

\begin{figure}[ht!]
\plotone{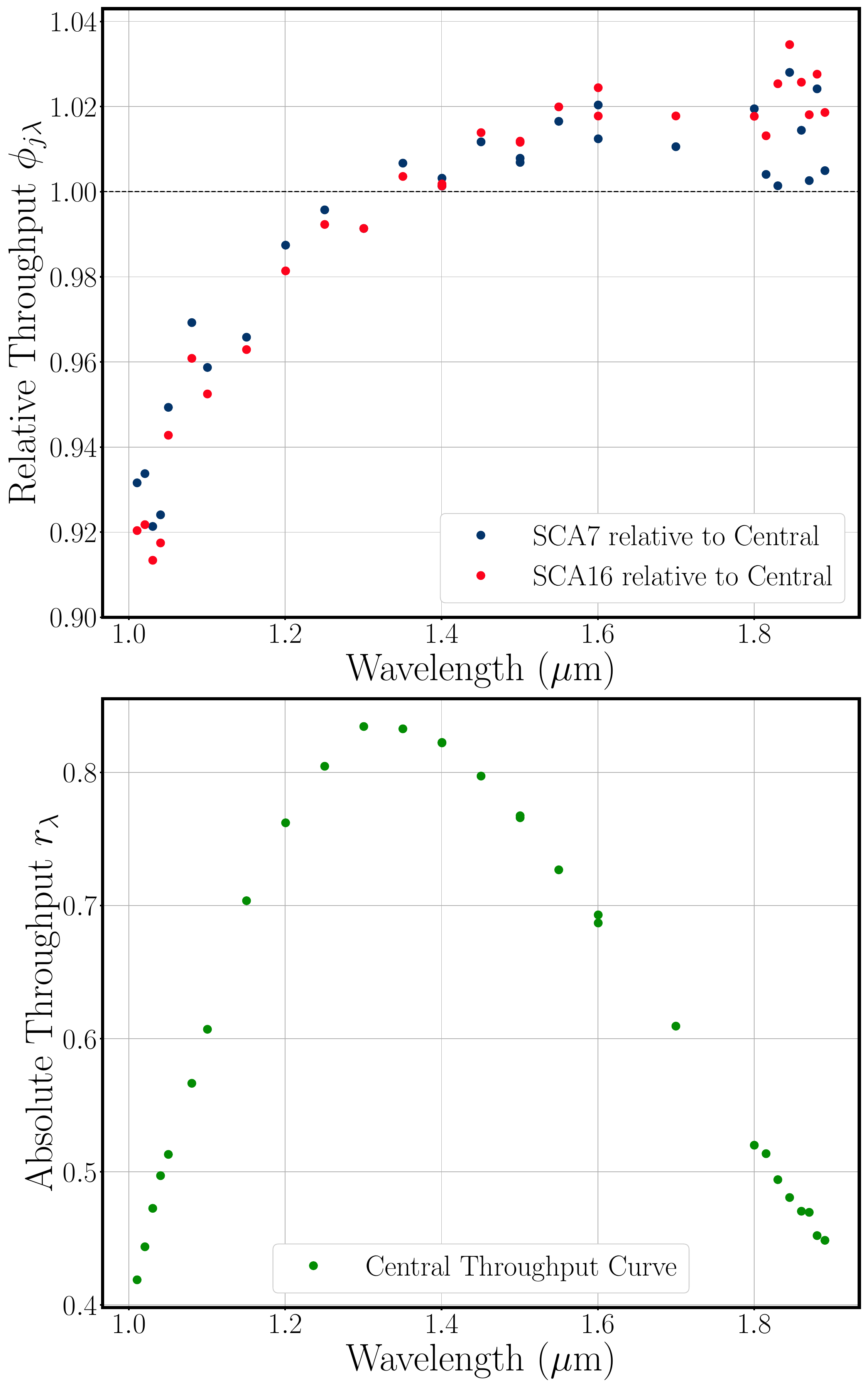}
\caption{The left panel shows the throughput of the central pixel of SCA7 and SCA16 (coloured blue and red, respectively) relative to the WFI central point. The right panel shows the absolute throughput curve measured at the WFI centre. The positions of these measurements are shown in Figure~\ref{fig:throughput_positions} with the same colouring.\label{fig:throughput}}
\end{figure}

\begin{figure}[ht!]
\plotone{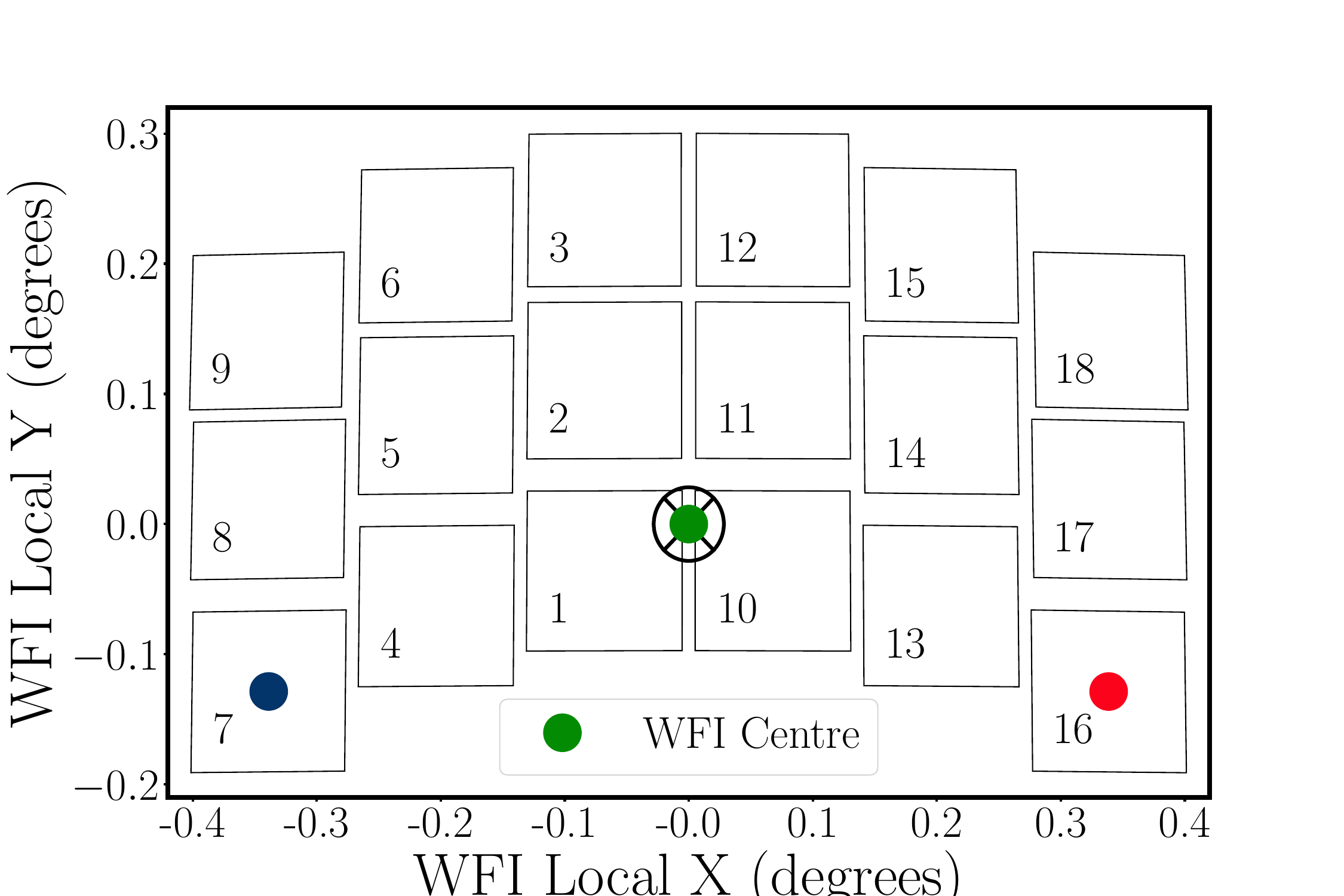}
\caption{Positions of throughput measurements shown in Figure~\ref{fig:throughput}.\label{fig:throughput_positions}}
\end{figure}

\begin{figure}[ht!]
\plotone{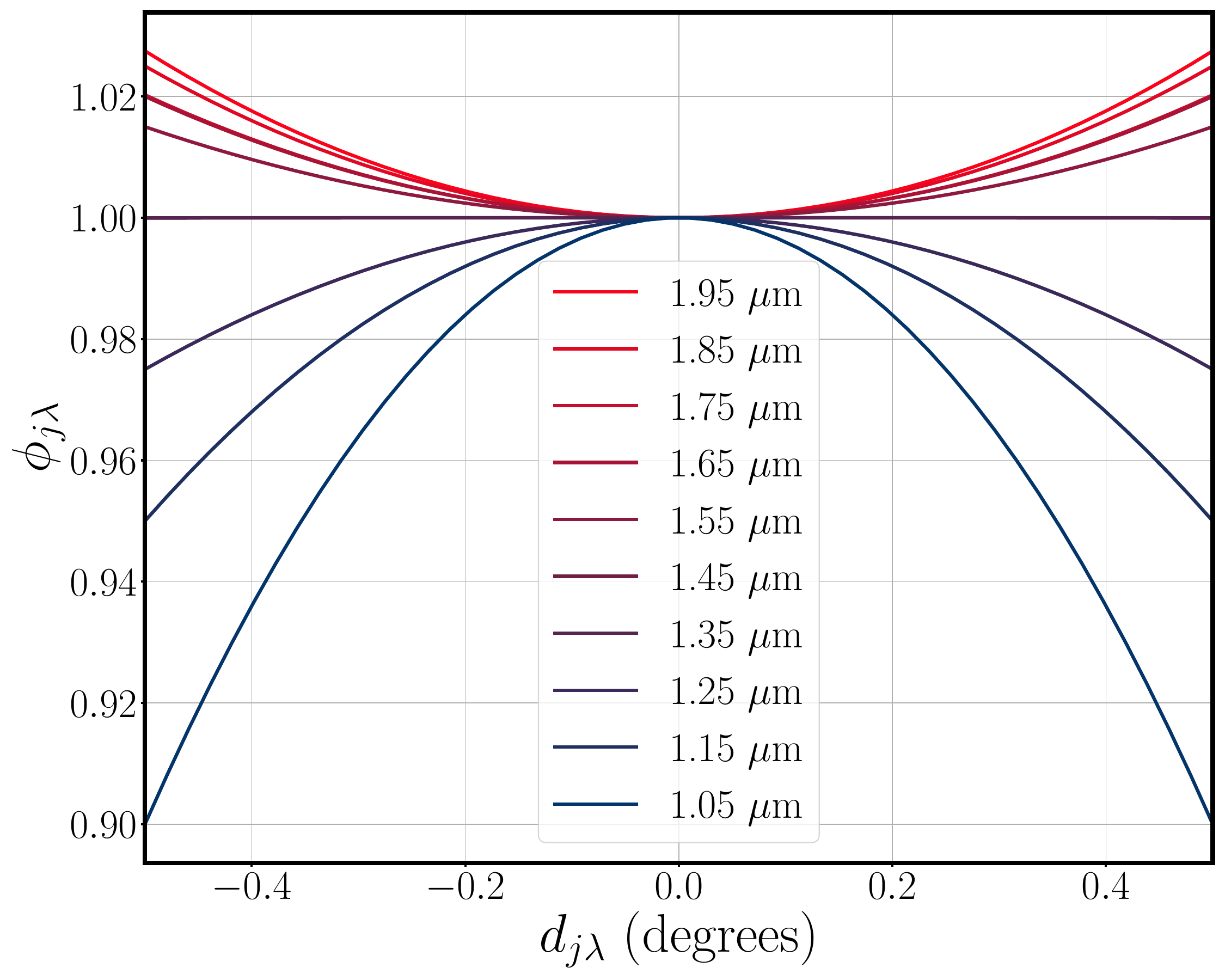}
\caption{Slices of the paraboloid shape of the throughput. The paraboloid is symmetric around the centre of the focal plane and is only a function of the distance from the WFI centre. \label{fig:paraboloid}}
\end{figure}

Given a set of wavelength and focal plane position dependent throughputs, we construct a simple model for the focal plane variation calibration. From \cite{roman_throughput_2024}, we have measurements of the throughput curve (lower panel of Figure~\ref{fig:throughput}) as a function of wavelength at the three positions shown in Figure~\ref{fig:throughput_positions}. The upper panel of Figure~\ref{fig:throughput} shows the symmetric drop off of the relative throughput at low wavelengths, and the slight increase in throughput at higher wavelengths. Thus, we construct a simple paraboloid model to capture this effect. This effect is only a function of the wavelength and position of the calibrator on the focal plane. A slice of these parabolic throughputs are shown in Figure~\ref{fig:paraboloid}. The $A_\lambda$ values for each wavelength are shown in Table~\ref{tab:alam}.

\begin{table}[h]
\centering
\caption{Values of $A_\lambda$ and corresponding wavelength bin centres. Bins are 0.1 $\mu$m in width.}
\begin{tabular}{c|c}
\hline
Bin ($\mu$m) & $A_\lambda$ \\
\hline
\hline
1.05 & $-$0.11 \\
1.15 & $-$0.10 \\
1.25 & $-$0.081 \\
1.35 & $-$0.080 \\
1.45 & $-$0.060 \\
1.55 & 0.00020 \\
1.65 & 0.00010 \\
1.75 & 0.10 \\
1.85 & 0.20 \\
1.95 & 0.40 \\
\hline
\end{tabular}
\label{tab:alam}
\end{table}

\subsection{Modelling the HLSS}
\label{sec:hlwas_modelling}

\begin{figure}[ht!]
\plotone{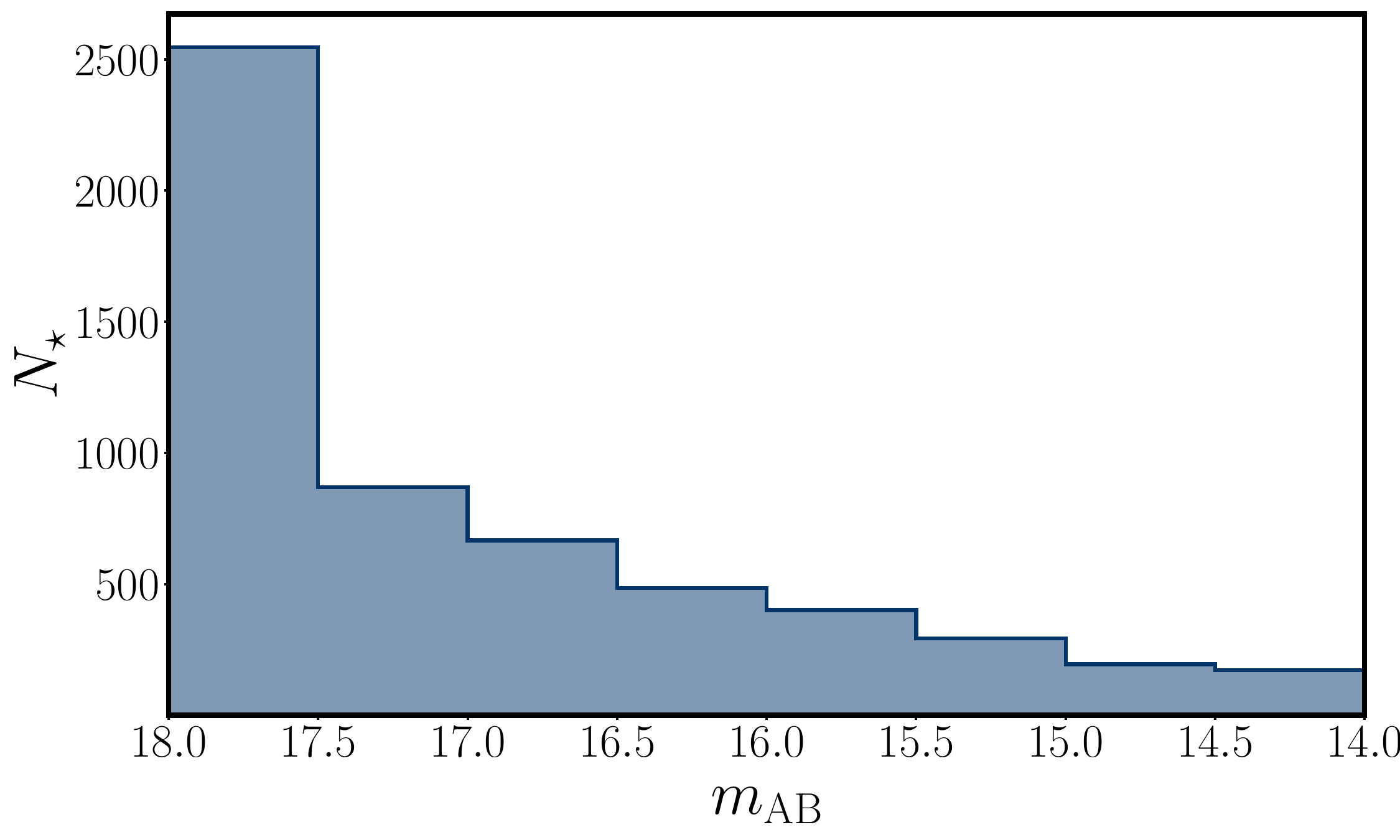}
\caption{Histogram of calibrator magnitude in a simulated four square degree patch of the Milky Way centred on RA = 10.0$^{\circ}$, DEC = 0.0$^{\circ}$. There are a total of 5631 calibrators. \label{fig:stardist}}
\end{figure}

We simulate a field of point like calibrators over the bolometric magnitude range of interest in a four square degree field of view between centred at RA = 10$^{\circ}$ and DEC = 0$^{\circ}$ (Shown in Figure~\ref{fig:examplesurvey}). We populate the region with calibrators from magnitude 14 to 18 (over the same wavelength range as above) and estimate measurement errors for all wavelength bins and magnitudes using Pandeia with the latest parameters for the HLSS \citep{pandeia, roman_rotac_2025}.\footnote{As of writing, a 195 second exposure with the $\rm SP\_190\_12$ MultiAccum table \citep{WFI_MultiAccum_Tables_2022}.} The distribution of calibrators in this patch is shown in Figure~\ref{fig:stardist}.\footnote{This distribution was generated by the same code used for \cite{wang_high_2022}, based on \cite{star_dist1_chang2010, star_dist2_just2015, star_dist3_juric2008}.}

The exposures will be dithered to cover parts of the survey that would fall into the chip gaps. In addition, each dither will have four rolls to change the angles the wavelength traces appear on the detectors. This primarily helps to alleviate the issue of spectrum confusion. This survey pattern will result in regions of the sky with between zero and eight survey passes. This paper will not consider maximising the high pass coverage. 

We will consider a fixed set of eight points and only vary the dither between the first set of four rolls and the second four rolls. We generate an initial exposure at a given pointing, then simulate three additional rolls at the same pointing. Next, we shift the RA and DEC of the pointing by a small amount $d_{\rm RA} = \frac{1}{2}d_{\rm DEC}$, and repeat the process to obtain the chunk for the given pointing. This is repeated for all eight pointings to give a mask for our simulated survey. The SCA positions are used to determine which calibrators in the field are ``observed" by which detector exposures, assigning each calibrator to one or more exposures. This generates a matrix $m_{ij}$ which is non-zero where the $j^{\rm th}$ calibrator is observed in the $i^{\rm th}$ detector exposure.

\subsection{Grism Information}
\label{sec:grism}

Given the survey geometry and calibrator sky positions, we now find the sky positions of the wavelength traces. The calibrator sky coordinates are transformed into the detector frame, then the Roman optical model is applied to obtain the wavelength trace in detector coordinates. The trace is then binned into ten evenly spaced wavelength bins (See Table~\ref{tab:alam}), giving mean detector coordinates for each bin. These coordinates are then transformed back into sky coordinates. We only consider the first order wavelength traces and assume no trace confusion. As these calibrators are already assigned to a detector, we then may simply extend our $m_{ij}$ matrix into the $\lambda$ dimension. In cases where a trace spans multiple detectors, we only keep the portion that is on the same detector as the actual calibrator position. Once the three effects outlined in Section~\ref{sec:obs_sim} are applied we have finally completed generating our measurement matrix $m_{ij\lambda}$.

The transformations between the telescope and sky coordinates were handled with the package \texttt{PYSIAF} \citep{pysiaf}. 

\section{Likelihood Maximisation}
\label{sec:likemax}

We can start by constructing an effective $\chi^2$, solving for both the calibration parameters and the true stellar brightnesses per wavelength bin $m^\star_{j\lambda}$, including a prior term on the $\hat{k}_i$:

\begin{align}
\chi^2_{\mathrm{eff}}(\hat{k}_i, \hat{A}_{\lambda}, \hat{r}_{\lambda} | m_{ij\lambda}) &= 
\sum_{i,j,\lambda} \frac{(m^c_{ij\lambda} - m^\star_{j\lambda})^2}{\sigma_{j\lambda}^2} \nonumber \\
&+ \sum_i^{N^{\mathrm{exp}}} \left( \frac{\hat{k}_i}{\sigma_k} \right)^2 ,
\label{eq:chi2_full}
\end{align}

\noindent
where $\sigma_{j\lambda}$ is the measurement error of the $j^{\rm th}$ calibrator observed in the $\lambda^{\rm th}$ wavelength bin and $\sigma_k$ is the width of the assumed Gaussian distribution of the independent $k_i$, centred on zero. $N^{\rm exp}$ is the total number of detector exposures.

We undertake the optimisation in two steps. First, we average $m^c_{ij\lambda}$, the ``calibrated" measurement matrix with $\hat{k}_i, \hat{A}_{\lambda}, \hat{r}_{\lambda}$ applied, along the $i$ axis, forming an estimate of the true brightness of each calibrator in each wavelength bin. $N^{\rm exp}_j$ is the number of observations of the $j^{\rm th}$ calibrator:\footnote{If $\hat{k}_i$, $\hat{r}_{\lambda}$, $\hat{\phi}_{j\lambda}$ are perfectly recovered, then clearly $\overline{m}^c_{j\lambda} = m^\star_{j\lambda}$.}
\begin{align}
    \overline{m}^c_{j\lambda} &= \frac{1}{N^{\rm exp}_j}\sum_{i}m^c_{ij\lambda} \nonumber \\ &= \frac{1}{N^{\rm exp}_j}\sum_{i}m_{ij\lambda}+ \hat{k}_i + \hat{r}_{\lambda} + \hat{\phi}_{j\lambda}.
    \label{eq:mcjlam}
\end{align}

When inserting this estimate for the true brightnesses into Equation~\ref{eq:chi2_full},  $r_\lambda$ cleanly cancels out of the likelihood and we are left with the following equation that recovers the set of $\hat{k}_i$ and $\hat{A}_\lambda$,
\begin{align}
&\chi^2_{\mathrm{eff}}(\hat{k}_i, \hat{A}_{\lambda}| m_{ij\lambda}) \nonumber \\
&= \sum_{i,j,\lambda} \frac{(m_{ij\lambda} + \hat{k}_i + \hat{\phi}_{j\lambda} - \frac{1}{N^{\rm exp}_j}\sum_{i}[m_{ij\lambda} + \hat{k}_i + \hat{\phi}_{j\lambda}])^2}{\sigma_{j\lambda}^2} \nonumber
\\&+ \sum_i^{N^{\mathrm{exp}}} \left( \frac{\hat{k}_i}{\sigma_k} \right)^2.
\label{eq:chi2_kA}
\end{align}

We note here that either the assumption of flat spectra, or knowledge of the shape of the spectral shape is necessary for the second step as the spectral shape is degenerate with the wavelength throughput. However, the first step does not have this requirement, thus in the event of poor knowledge of the calibrator spectra, the $\hat{k}_i$ and $\hat{A}_\lambda$ may be recovered, and we would leave the $r_{\lambda}$ to the absolute calibration to handle.

Having recovered the $\hat{k}_i$ and $\hat{A}_\lambda$, we apply them to the measurement matrix. The only remaining effect is the wavelength dependent throughput $\hat{r}_\lambda$. We then form a second estimate of the true bolometric brightness of the calibrators in each exposure, $N^{\rm bins}_j$ is the number of wavelength bins the $j^{\rm th}$ calibrator is observed in:
\begin{equation}
        \overline{m}^c_{ij} = \frac{1}{N^{\rm bins}}\sum_{\lambda}m_{ij\lambda} + \hat{k}_i + \hat{r}_{\lambda} + \hat{\phi}_{j\lambda},
    \label{eq:mcij}
\end{equation}

\noindent
where we now have a good estimate of $\hat{k}_i$ and $\hat{\phi}_{ij\lambda}$. Continuing with the second $\chi^2$:
\begin{equation}
\chi^2_{\mathrm{eff}}(\hat{r}_{\lambda}| m_{ij\lambda}, \hat{k}_i, \hat{A}_\lambda) = 
\sum_{i,j,\lambda} \frac{(m^c_{ij\lambda} - \overline{m}^c_{ij})^2}{\sigma_{j\lambda}^2},
\label{eq:chi2_r}
\end{equation}
\noindent
and we recover the $\hat{r}_\lambda$. After taking the best estimate of the $\hat{r}_\lambda$ and applying this second correction, we compute the calibration residuals $\varepsilon_{ij\lambda}$ (Equation~\ref{eq:residuals}) and the standard deviation of the nonzero values serves as our quality metric, $\sigma_f$.

It is also possible to directly recover all the calibration parameters at once by minimising the following form of the $\chi^2$:
\begin{align}
\chi^2_{\mathrm{eff}}(\hat{k}_i, \hat{A}_{\lambda}, \hat{r}_{\lambda} | m_{ij\lambda}) &= 
\sum_{i,j,\lambda} \frac{(m^c_{ij\lambda} - \overline{m}^c_{j})^2}{\sigma_{j\lambda}^2} \nonumber \\
&+ \sum_i^{N^{\mathrm{exp}}} \left( \frac{\hat{k}_i}{\sigma_k} \right)^2.
\label{eq:chi2_full_mean}
\end{align}

$\overline{m}^c_{j}$ is given by a mean over both exposures $i$ and wavelength bins $\lambda$:
\begin{align}
    \overline{m}^c_{j} &= \frac{1}{N^{\rm exp}_jN^{\rm bins}_j}\sum_{i,\lambda}m^c_{ij\lambda} \nonumber\\ &= \frac{1}{N^{\rm exp}_jN^{\rm bins}_j}\sum_{i,\lambda}m_{ij\lambda} + \hat{k}_i + \hat{r}_{\lambda} + \hat{\phi}_{j\lambda}.
    \label{eq:mcj}
\end{align}
\noindent
However, a slight degeneracy is introduced when recovering the $\hat{k}_i$ and $\hat{r}_\lambda$ parameters in the $\overline{m}^c_{j}$ term. This both degrades the recovery of these parameters and significantly slows the convergence of the minimiser. 

\section{\textit{{\"U}bercalibration} Results}
\label{sec:ubercalresults}

\begin{figure*}[ht!]
\plotone{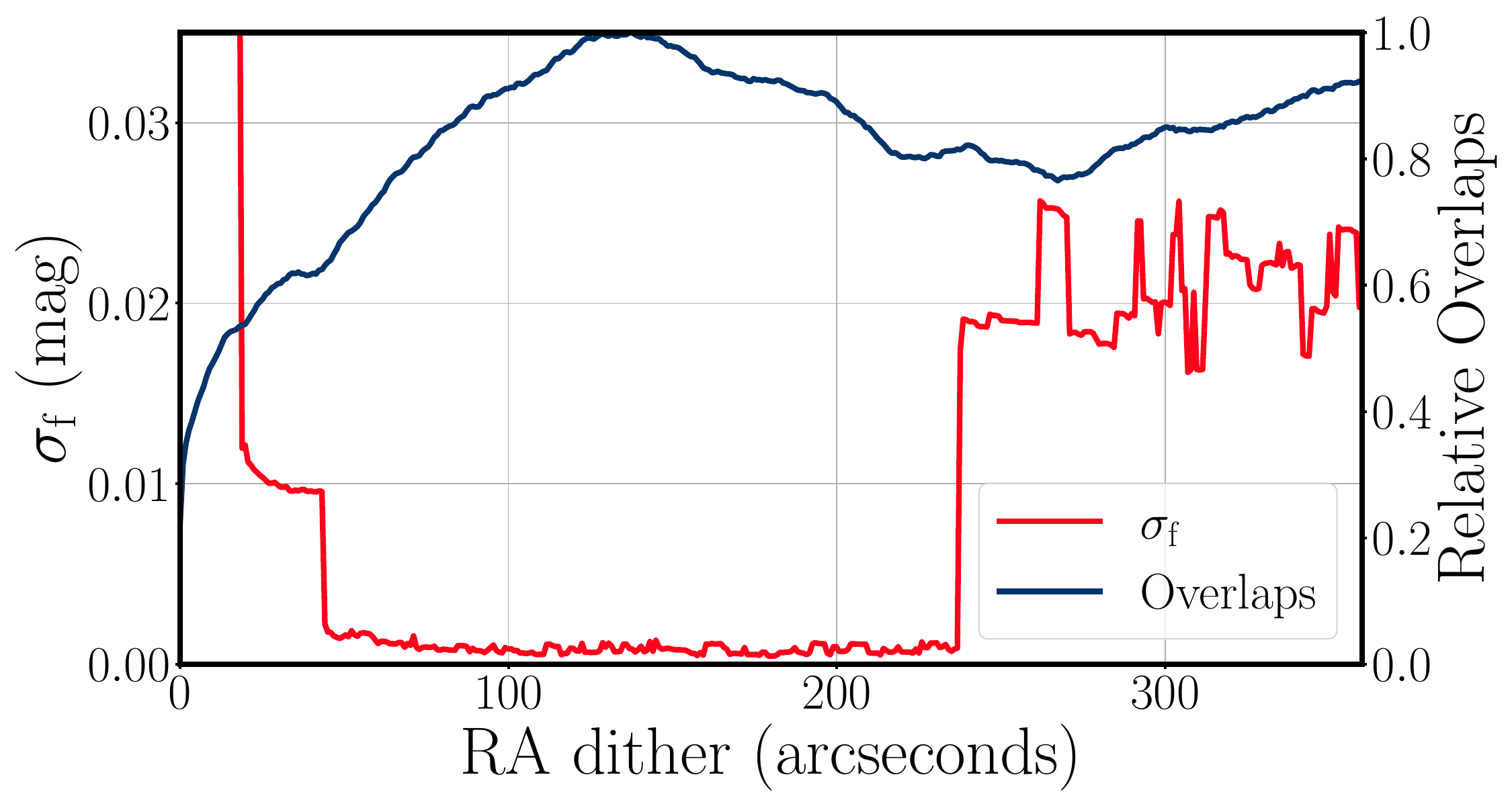}
\caption{$d_{\rm RA}$ scaling of the final zero point scatter. In this figure, we use the reference magnitude cut between 14 and 17 magnitude. The left vertical axis shows the final residual distribution width in magnitudes. The right vertical axis scales the number of overlaps relative to the maximum at about 140$\arcsec$. \label{fig:ditherscaling}}
\end{figure*}

Utilising the simulated survey and the simulated star field, we input a flat field calibration for each detector exposure, the constant wavelength dependent response, and the smoothing varying field of view throughput, generating the uncalibrated measurement matrix $m_{ij\lambda}$. We then use the L-BFGS-B algorithm from the \texttt{scipy.minimize} library to find the sets of $\{\hat{k}_i\}$ and $\{\hat{A}_\lambda\}$ that minimise Equation~\ref{eq:chi2_kA}, then use the same algorithm to find the set of $\{\hat{r}_\lambda\}$ that minimises Equation~\ref{eq:chi2_r}. We assume there is no coherent time variation of the flat field calibrations. We show the results of this procedure on a catalogue selected between 14 and 17 magnitude in Figure~\ref{fig:ditherscaling}.

We find the minimum $\sigma_{\rm f} = 0.44 \ \rm mmag$ with $d_{\rm RA} =179.1\arcsec$. We see the width drops rapidly as the dither increases from zero, with the first significant feature around $d_{\rm RA}$ $\approx$ 40$^{\prime\prime}$. This where the dither first covers the smallest horizontal chip gaps in the WFI focal plane (See Figure~\ref{fig:throughput_positions}). In this range, the power of the calibration is primarily driven by the number of unique overlaps in the calibration. We define an overlap to be a region of the survey that is observed by multiple exposures.\footnote{For example, given four unique exposures, there is a maximum of 12 possible unique overlaps (ignoring single exposure regions).} In Figure~\ref{fig:ditherscaling}, we see the peak of the number of overlaps corresponds to the approximate minimum in $\sigma_{\rm f}$ Then, we see a feature at 240$\arcsec$. This represents an important scale in the calibration. This is approximately half an SCA width and half the maximum vertical chip gap ($d_{\rm dec}$ = 2$d_{\rm RA}$). It is at this point that the survey mask becomes less interconnected as this dither introduces a perfect vertical alignment of the detectors. This drives a minimum in the number of overlaps at this point. While the number of overlaps begins to climb after this scale, the disconnect between detectors dominates this region. Thus, the optimal dither is between these two scales, which gives a large possible range. It is important to note that a larger dither would require more telescope resources. We leave it to future work to correctly balance this. Recalling the goal of a 1$\%$ flux calibration (equivalent to 10.0 mmag), we see that a large range of dithers is viable for this goal, giving significant flexibility to other pieces of the observing program to be optimised without strongly affecting the effectiveness of the relative flux calibration procedure.

\begin{figure}[ht!]
\plotone{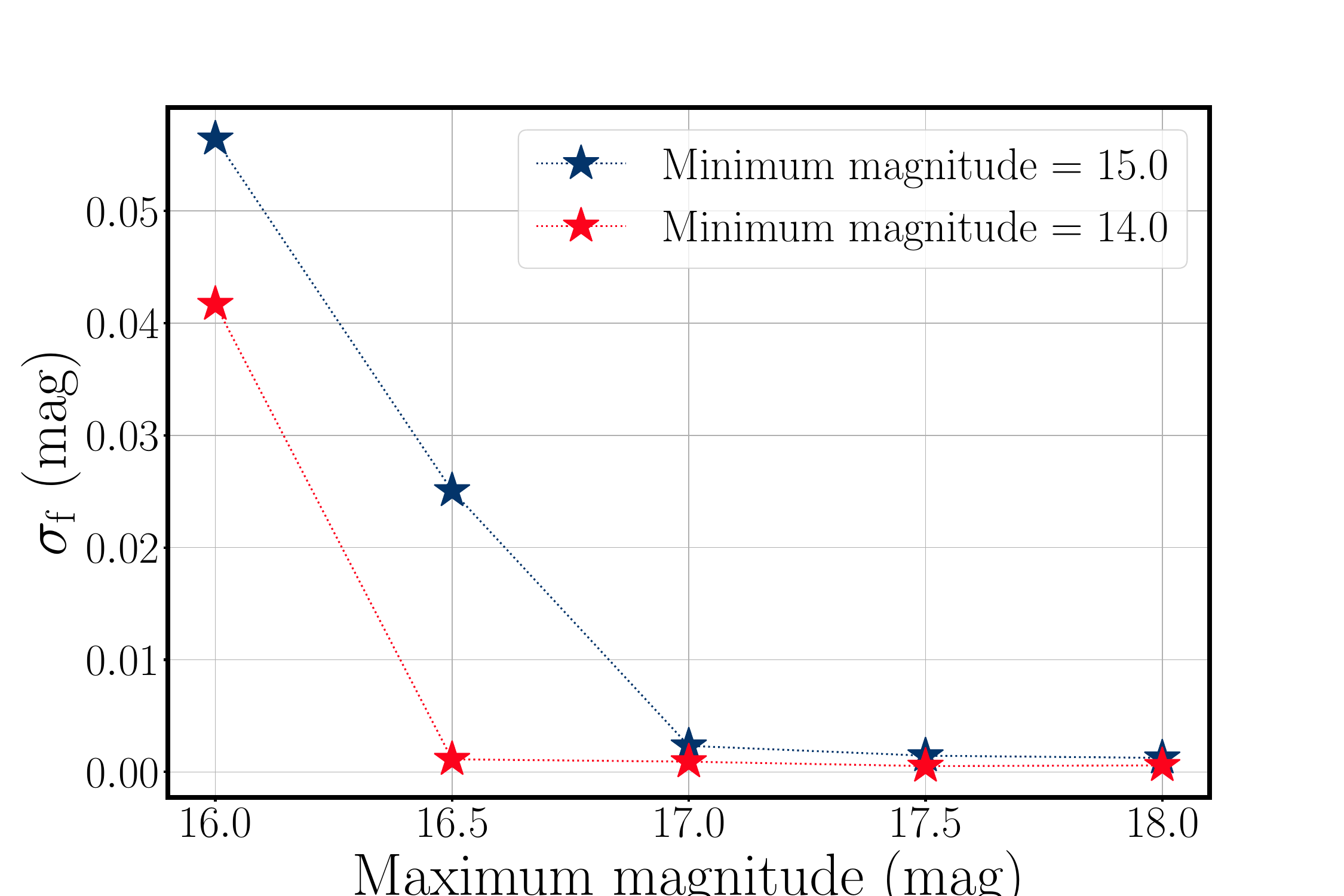}
\caption{Scaling of the final zero point scatter as a function of the magnitude cut of the calibrator population. In this figure, we use the reference $d_{\rm RA} = 90"$.\label{fig:magnitudescaling}}
\end{figure}

We also consider the selection of calibrator magnitude for the relative flux calibration. We consider using bright magnitude cuts at 14 and 15 magnitude and faint cuts between 16 and 18 magnitude. Here we use a reference dither of $d_{\rm RA}=$ 90$\arcsec$. These results are shown in Figure~\ref{fig:magnitudescaling}.


We see the exclusion of the 14 and 14.5 magnitude calibrators significantly degrades the power of the calibration. While there are relatively few calibrators in this magnitude range, their measurement error is small, allowing us to tie down the calibration with less uncertainty. This is not to say they must be included. If we assume a similar dither scaling, the 1$\%$ calibration goal should be easily achievable using a magnitude range starting at 15 provided the maximum cutoff is more than 17 magnitude. This may be necessary to prevent saturation effects due to the overwhelming brightness of these calibrators. In addition, we note that the including calibrators past 17 magnitude does not significantly improve the calibration power.

\section{Discussion}
\label{sec:discussion}

In this section, we discuss some additional steps that should be strongly considered prior to applying the calibration model to real Roman data.

Some immediate considerations are the assumptions made regarding the covariance of the various components of the calibration model. While it remains reasonable that the calibrator brightnesses be independent, the flat fields of adjacent detectors are unlikely to be fully independent, due to the geometry of the optics. While this effect may be somewhat mitigated by the inclusion of the parabolic (or more complicated) focal plane throughput model, it remains to be seen how well this issue can be constrained only at the data level. 

We have applied a simple model given our limited data to the wavelength dependent calibration. The model adopted for the multiplicative focal plane response was a paraboloid of the form $A d^2 + 1$. In practice, the true shape of the calibration may be arbitrarily complex, and thus difficult to characterise prior to launch. To mitigate this lack of knowledge, we may consider generalising the focal plane response to be composed of functions of some arbitrary basis. A generalisation of the focal plane response has been explored in \cite{general_calibration_davini_2021}. We can take our response as a special case of a polynomial basis cut off at the second degree. However, we note that given the current two step implementation of calibration, the number of basis functions used for each wavelength bin must either be the same or predetermined. We do not see any reason for this to cause issues, as there is not a strong reason for the overall shape of the response to change as a function of wavelength.

Such a polynomial basis was applied to the Gaia DR1 large scale calibration, considering a second degree cutoff in the focal plane position analogue and up to a 6th degree cutoff in the colour calibration. While the model for the calibration was not significantly more complex in form than the one explored in this paper, calibration parameters were recovered for a large number  of calibration units ($\sim 10^6$): combinations of observing parameters such as colour, CCD, pixel, and observation time \citep{gaia_dr1_2016, gaia_dr1_cal_carrasco_2015}. One effect to highlight is the variation of telescope sensitivity over time, making it so that any relative calibration is only valid for certain periods of the survey, depending on the timescale of these effects. While it may not be feasible (or time efficient) to undertake a full relative flux calibration over entire survey passes, we can consider a two stage \textit{{\"u}bercalibration}. In \cite{survey_design_holmes_2021}, it is shown that the \textit{{\"u}bercalibration} is improved either by the inclusion of more calibrators or more exposures of the same calibrators (at different overlaps). Furthermore, we saw above that the inclusion of calibrators with low brightness uncertainty vastly improves the final calibration. Thus, we propose a complex model accounting for pixel-to-pixel fluctuations to be constrained using the HLWAS deep fields. Over the nominal survey, these deep fields are to be returned to ten times, also allowing for the tracking of the fine calibrations over the survey lifetime. Then, we may assume that any changes to the calibration are small perturbations to the model obtained from the deep field calibration, and apply a much simpler model over the wider footprint.

As mentioned in Section~\ref{sec:likemax}, the assumption of flat spectra in our analysis is unrealistic. With real calibrator stars, the non-trivial shape of their spectra is degenerate with the overall wavelength dependent throughput. There are two possibilities that can address this. First, if we have measurements of the shape of the calibrator spectra, then it is straightforward to include that information into the likelihood maximisation. In this case, the problem reduces to a flat spectrum assumption. Second, if the spectral shapes of the calibrators are poorly constrained, we could leave the wavelength constraint to the overall absolute calibration. In this case, we would apply the \textit{{\"u}bercalibration} separately to each wavelength bin, constraining geometric effects. With a relative calibration in each wavelength bin, an absolute wavelength calibration can be applied to correctly set the wavelength bins to a common absolute scale.

Finally, we make note of the significant amount of ground testing that the WFI has undergone and will continue to undergo before the launch date. This testing will provide the best idea of how Roman performs in the vacuum of space while still on the ground. As of writing, the ground testing campaign is completed, from the characterisation of individual WFI components like the SCAs, and filters, to the testing of the entire WFI assembly in thermal vacuum. Only data analysis remains for the final testing phase of installed WFI on the scientific payload \citep{ground_testing_roman}. Due to the extremely stringent calibration requirements of Roman, further calibration will be required when on sky post-launch. Divided into three parts, these are internal calibrations, external calibrations, and advanced calibrations (such as the \textit{{\"u}bercalibration} explored in this paper). The internal calibration will be carried out using onboard light sources to determine standards like the gain, dark, and flat fields. Conversely, the external calibration is meant to be carried out through repeated observations of a set of dedicated touchstone fields. In particular, the grism will visit two dedicated fields with a monthly cadence. Of these, one will be for the spectrophotometric relative flux calibration. As of writing, the observations of this field will be significantly deeper than those explored in this paper, with an expected minimum of 20 dithers at 300 seconds of exposure \citep{touchstone_report}. We have explored the possibility of exploiting the regularity of these fields to improve our relative flux calibrations, however these fields provide significant value beyond the scope of this paper, such as the opportunity for characterisation of the point spread function over the survey lifetime. The current on sky calibration plan dedicates approximately six percent of the mission time to calibration \citep{sky_testing_roman}.

\section{Conclusions}
\label{sec:conclusions}

In this work, we have explored the application of an \textit{{\"u}bercalibration} style calibration to a survey like the spectroscopic component of the Roman HLWAS. We tested the effects of the dither size within a chunk, as well as the effect of the magnitude cuts on the calibrator field used for the calibration. To increase the speed of calculation and avoiding finding the true stellar magnitudes, we adapt a method introduced by \cite{markovic_large-scale_2017}.

We conducted our analysis on a simplified and smaller version of the Roman HLSS confined with a four square degree patch of sky with only point like calibrators and flat spectra. We defined three effects on the calibrators: a flat field calibration drawn from a Gaussian for each detector exposure, a constant wavelength dependent throughput, and a smoothly varying effect dependent on the calibrators position on the focal plane. Applying this model to the simplified survey, we find for the reference magnitude cut, the best zero point uncertainty is $\sigma_f = 0.44 \ \rm mmag$ with $d_{\rm RA} = 179.1\arcsec$, but values limited below 1.5 mmag are available in the RA dither range of 50-240$\arcsec$. While the actual numbers presented here should not be entirely taken at face value, dithers in this region both cover the chip gaps as required, without compromising the interconnectedness of the survey. We also note a near two to three fold improvement in the calibration comparing the calibration power between the edges of the optimal region and centre.

We have presented a study on the large scale relative spectrophotometric flux calibration of the Roman HLWAS. We have shown that this calibration is robust to the selection of the calibrator magnitudes, as well as the dither size, provided the dither is within a large ideal range. As the clustering measurements in the Roman scientific program strongly depend on an accurate photometric calibration, it is important that the survey geometries explored in this paper are taken into account when designing the overall observing program. However, we have shown that the ideal ranges are wide enough that any other survey requirements should be able to be comfortably met while maintaining a relative flux calibration well within the survey requirements.

\begin{acknowledgments}
All authors thank the members of the Roman Galaxy Redshift Survey Project Infrastructure Team for useful discussions. NP thanks Jeff Kruk for discussions on earlier results. We gratefully acknowledge funding from NASA Grant $\#$80NSSC24M0021, ``Project Infrastructure for the Roman Galaxy Redshift Survey''.

\end{acknowledgments}

\appendix
\section{Table of Symbols}
\label{sec:gloss}

In this appendix, we summarise the symbols and abbreviations used in this paper for the reader's convenience.

\begin{table*}[!ht]
\centering
\begin{tabular}{l|l|r}
\hline
\textbf{Symbol} & \textbf{Description} & \textbf{Unit} \\
\hline
Roman & Nancy Grace Roman Space Telescope & None \\
HLWAS & The High Latitude Wide Area Survey on Roman & None \\
HLSS & Spectroscopic component of the HLWAS on Roman & None \\
WFI & Wide Field Instrument & None \\
SCA & Sensor Chip Assembly & None \\
CCD & Charge Coupled Device & None \\
NIR & Near Infrared & None \\
FoV & Field of View & None \\
RA & Right Ascension & deg \\
DEC & Declination & deg \\
\hline
$N^{\rm exp}$ & Number of detector exposures. 1152 in this work & None \\
$N_{\star}$ & Number of calibrators ``observed''. Depends on dithering & None \\
$N^{\rm bins}$ & Number of wavelength bins. 10 in this work & None \\
\hline
i & indexes detector exposures between 1 and $N^{\rm exp}$ & None \\
j & indexes calibrators between 1 and $N_{\star}$ & None \\
$\lambda$ & indexes wavelength bins between 1 and $N^{\rm bins}$ & None\\
\hline
$N^{\rm exp}_j$ & Number of observations of the $j^{\rm th}$ calibrator & None \\
$N^{\rm bins}_j$ & Number of wavelength bins the $j^{\rm th}$ calibrator is observed in & None \\
\hline
$k_i$ & True flat field offset & mag \\
$\sigma_k$ & Width of assumed Gaussian distribution of $k_i$ & mag \\
$r_\lambda$ & True wavelength throughput curve & mag \\
$\phi_{j\lambda}$ & True FoV throughput & mag \\
$A_{\lambda}$ & True quadratic coefficient for $\phi_{j\lambda}$ & $\rm deg^{-2}$ \\
$\hat{k}_i$ & Recovered flat field offset & mag \\
$\hat{r}_\lambda$ & Recovered wavelength throughput curve & mag \\
$\hat{\phi}_{j\lambda}$ & Recovered FoV throughput & mag \\
$\hat{A}_{\lambda}$ & Recovered quadratic coefficient for $\hat{\phi}_{j\lambda}$ & $\rm deg^{-2}$ \\
\hline
$m^{\star}_{j\lambda}$ & True magnitude of $j^{\rm th}$ calibrator in the $\lambda^{\rm th}$ bin & mag \\
$m_{ij\lambda}$ & ``Observed'' magnitudes & mag \\
$e_{j\lambda}$ & Poisson error of the $j^{\rm th}$ calibrator in the $\lambda^{\rm th}$ bin & mag\\
$\sigma_{j\lambda}$ & Assumed width of the $e_{j\lambda}$ distribution & mag \\
$d_{j\lambda}$ & Distance from WFI centre of $\lambda^{\rm th}$ bin of the $j^{\rm th}$ calibrator & deg \\
$m^c_{ij\lambda}$ & Calibrated magnitudes & mag \\
$\overline{m}^c_{j\lambda}$ & Mean calibrated magnitude of the $j^{\rm th}$ calibrator in bin $\lambda$ & mag \\
$\overline{m}^c_{ij}$ & Mean calibrated magnitude of the $j^{\rm th}$ calibrator in detector exposure $i$ & mag \\
$\overline{m}^c_{j}$ & Mean calibrated magnitude of the $j^{\rm th}$ calibrator & mag \\
$\varepsilon_{ij\lambda}$ & Residual miscalibrations & mag \\
$\sigma_f$ & Width of the residual miscalibrations & mag \\
\hline
$d_{\rm RA}$ & Horizontal or RA dither distance & arcsec \\
$d_{\rm DEC}$ & Vertical or DEC dither distance & arcsec \\
\hline
\end{tabular}
\caption{Table of symbols and abbreviations used in this paper.}
\end{table*}

\section{A common failure mode}
\label{sec:failuremodes}
\begin{figure}[ht!]
\plotone{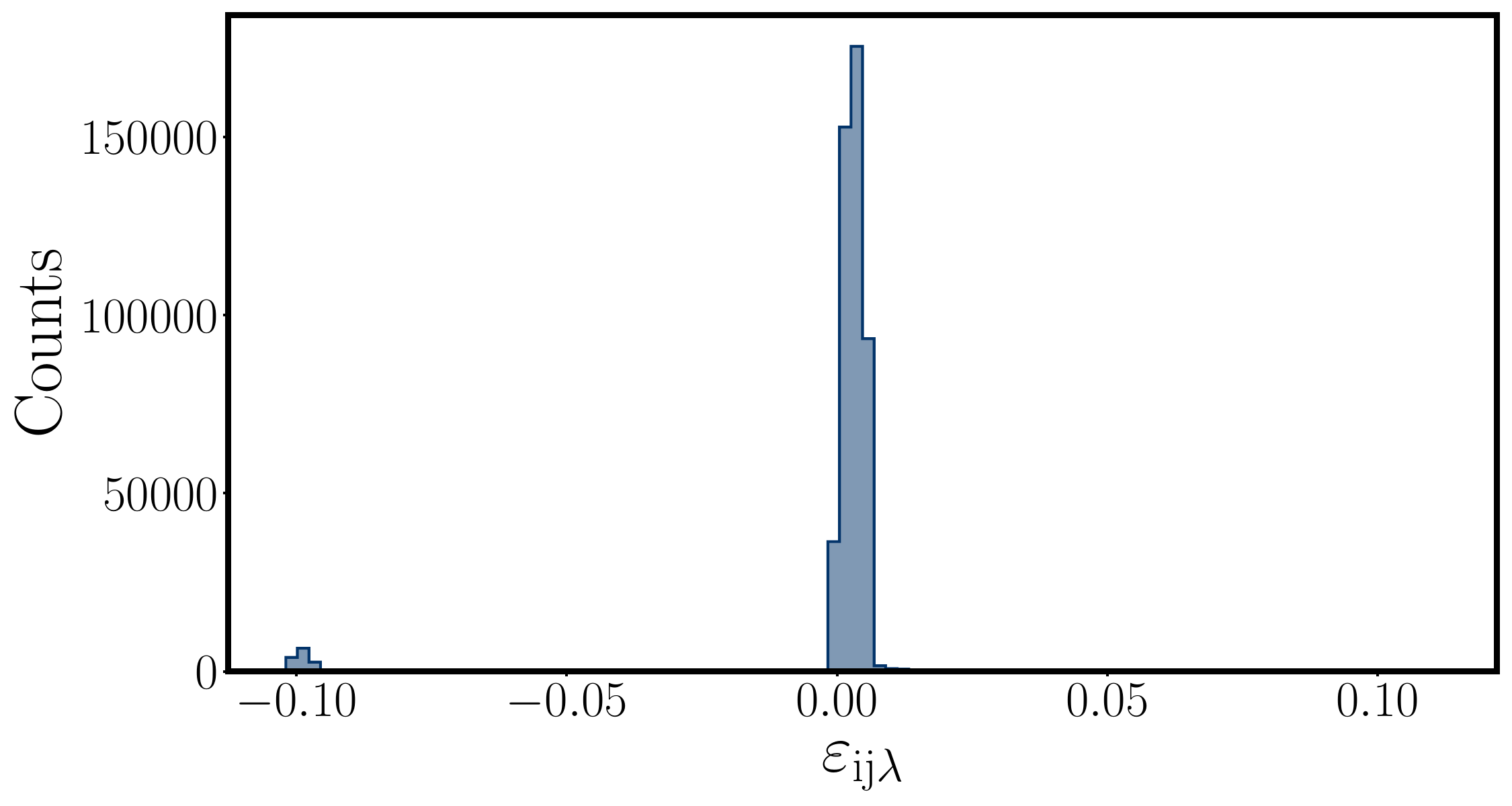}
\caption{Distribution of residuals for the method applied to a single chunk. Note the central Gaussian envelope with some failures at -0.10. \label{fig:residual_ex}}
\end{figure}
\begin{figure}[ht!]
\plotone{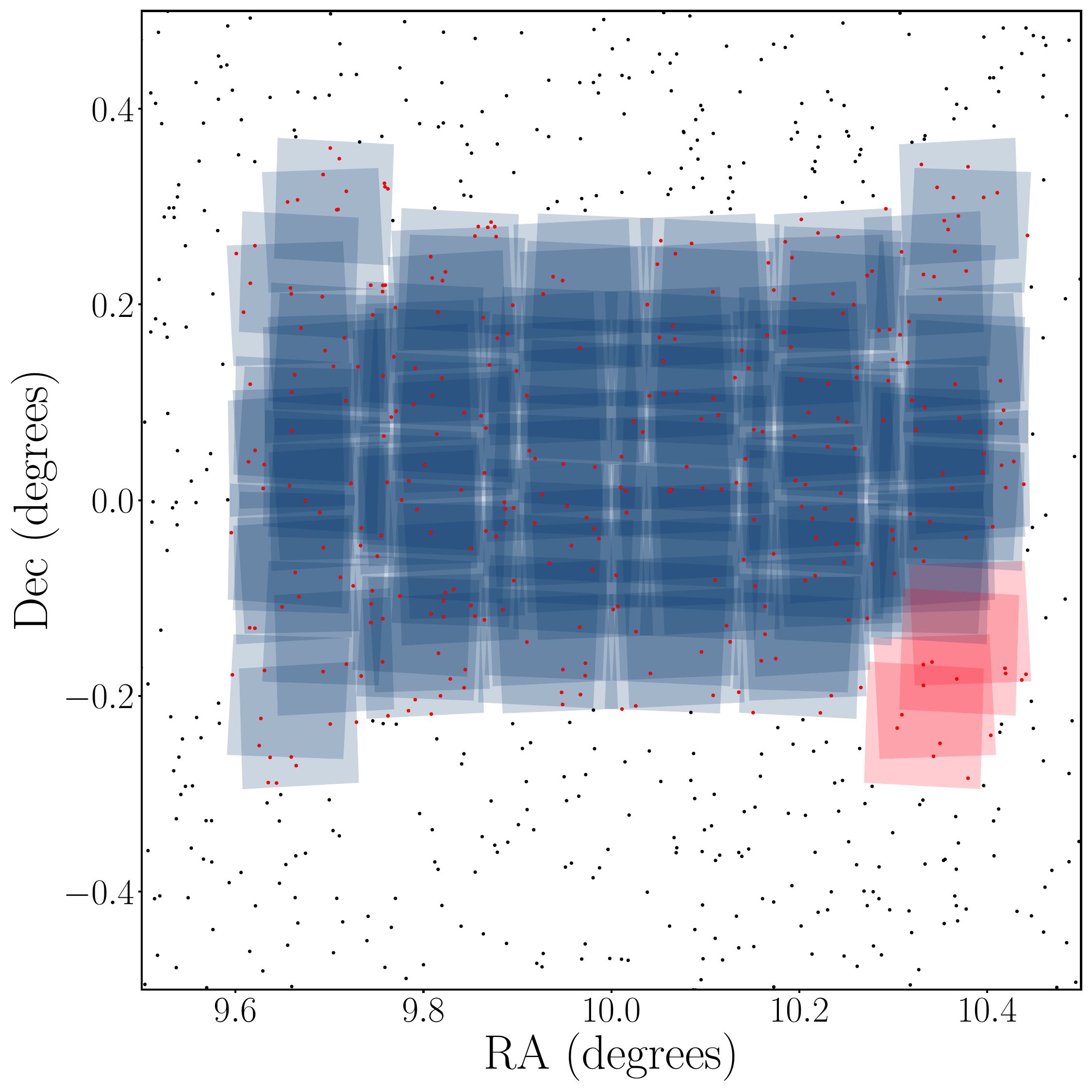}
\caption{Illustration of some failure modes. The outliers in the residual distribution (See Figure~\ref{fig:residual_ex}) are caused by the exposures highlighted in red. Unobserved calibrators are coloured black, while those that fall onto at least one exposure are coloured red. \label{fig:failuremodes}}
\end{figure}

In Figure~\ref{fig:residual_ex}, we show the distribution of residuals from applying the calibration method to a single chunk. There are two features to note. While there is a sharply peaked Gaussian distribution in the centre, there is a secondary peak around $\sim$ -0.10 mag. These calibration failures are caused by the detector exposures highlighted in red (Figure~\ref{fig:failuremodes}). These exposures are both relatively sparsely populated by calibrators and have relatively few overlaps with other exposures. Also note the mean of the central peak is shifted slightly right. This is an artifact of the effective $\chi^2$ which forces the mean of the distribution to be zero. Due to the secondary peak, the central peak is shifted positively to enforce this mean. This systematic offset can be left to the absolute calibration.

\newpage
\bibliography{romancalibration}{}
\bibliographystyle{aasjournalv7}



\end{document}